\documentclass[%
aip,
jap,
reprint, 
floatfix,
noshowpacs,
noshowkeys,
preprintnumbers
]{revtex4-2}
\usepackage{mathptmx}
\usepackage{graphicx}
\usepackage{etoolbox}
\usepackage{dcolumn}
\usepackage{float}

\makeatletter
\def\@email#1#2{%
 \endgroup
 \patchcmd{\titleblock@produce}
  {\frontmatter@RRAPformat}
  {\frontmatter@RRAPformat{\produce@RRAP{*#1\href{mailto:#2}{#2}}}\frontmatter@RRAPformat}
  {}{}
}%

\def\maketitle{
\@author@finish
\title@column\titleblock@produce
\suppressfloats[t]}
\makeatother
\usepackage[english]{babel}
\newlength{\figsize}
\newlength{\widefigsize}
\newlength{\figsizethree}
\newlength{\subfigsize}
\newlength{\sketchfigsize}
\begin{document}

\title{Optical {Kerr} nonlinearity enhancement in high‑index metasurfaces via {Mie} void lattices}
\author{Andrey V. Panov}
\affiliation{
Institute of Automation and Control Processes,
Far Eastern Branch of Russian Academy of Sciences,
5, Radio st., Vladivostok, 690041, Russia}
\email{panov@iacp.dvo.ru}

\begin{abstract}
Recently, research in nanophotonics has turned toward Mie resonances in voids on the surface of high-refractive-index materials. 
The optical Kerr effect (OKE) in high-index membrane metasurfaces with Mie void lattices is investigated using three-dimensional finite-difference time-domain (FDTD) simulations, with gallium phosphide (GaP) as a model material. 
The effective nonlinear refractive index is extracted for empty spherical and truncated-cone (frustum) voids in a high-index slab.
Metasurfaces with isolated Mie void resonances yield only modest effective OKE enhancement, up to a factor of ten relative to bulk GaP. 
Mie void resonances in GaP metasurfaces are observable when the separation between voids exceeds approximately 220 nm; otherwise, modes in the high-index material between the voids prevail.
A much stronger response arises from the later modes developing in the high-index regions between closely spaced voids.
While the nonlinear figure of merit of Mie-void metasurfaces is limited for applications relying solely on energy-density enhancement, the open-cavity geometry offers advantages for hybrid systems that require access to the confined field, such as quantum emitters or nonlinear materials infiltrated into the voids.
\end{abstract}


\maketitle

\section{Introduction}

The emergence of \textit{Mie-tronics}---nanophotonics based on Mie scattering resonances in high-refractive-index dielectric nanostructures---has revolutionized light manipulation at subwavelength scales. 
While conventional dielectric Mie resonators confine electromagnetic energy within high-index materials to enhance both electric and magnetic light-matter interactions, a complementary paradigm has been recently proposed: Mie voids, where light is confined within air-filled cavities surrounded by high-index dielectrics \cite{Hentschel23}. 
This inverse geometry fundamentally alters light confinement physics, enabling resonant field enhancement in low-index regions while simultaneously mitigating material absorption losses \cite{Chen98}.

Mie voids operate on principles analogous to conventional Mie resonators but with reversed refractive index contrast. 
In this case, field confinement occurs predominantly in the void region rather than in the host dielectric material. 
The void can be filled by air or low-index liquids, such as water \cite{Kuznetsov25,Niinomi25} or propanol \cite{Arslan25}.
Recent experimental demonstrations have implemented this concept across multiple platforms:  on the top of silicon \cite{Hentschel23,Kuznetsov25,Fu26,Liao26,Goldberg26}, tungsten diselenide (WSe$_2$) \cite{Sarbajna25}, gallium arsenide (GaAs) \cite{Ludescher25,Arslan25,Arslan26}, van der Waals bismuth telluride (Bi$_2$Te$_3$) \cite{Lu26}, and silicon carbide (SiC) \cite{Maruyama25,Niinomi25}. 
These advances illustrate the ability of Mie voids to concentrate optical energy within accessible low-index volumes.

The field confinement of light emitters can be described by the Purcell factor, which quantifies the enhancement of the spontaneous emission rate in a nanostructure relative to that in free space.
The Purcell factor of Mie voids in a dielectric with refractive index $n=4$ reached 5 for broadband directional light sources \cite{Reichel26}.
Mie voids patterned on a silicon surface for quantum emission exhibited Purcell factors up to 2.6 \cite{Fu26}.
Another typical application of field confinement is the enhancement of optical nonlinearity.
Lu et al. proposed capping Mie void resonators patterned in a high-index substrate with a WS$_2$ monolayer \cite{Lu26}.
This led to a 25-fold enhancement of the second harmonic generation from WS$_2$ positioned on a resonant Mie void compared with that on an off-resonant void.

For Mie voids, which often appear as truncated cone-shaped depressions on a high-index surface, significant penetration of the electromagnetic field into the dielectric has been observed \cite{Niinomi25,Reichel26}. 
This behavior can be exploited to enhance nonlinear optical effects in high-index all-dielectric metasurfaces.
This work investigates one such nonlinear effect---the optical Kerr effect (OKE)---in high-index all-dielectric metasurfaces with Mie voids.
The optical Kerr effect, which manifests as an intensity-dependent refractive index change, represents a fundamental third-order nonlinear process with applications in ultrafast switching and modulation, optical limiting, and nonlinear sensing.

\section{Simulation details}

The effective intensity-dependent refractive index can be investigated numerically via three-dimensional finite-difference time-domain (FDTD) simulations of Gaussian beam propagation through a nonlinear optical sample~\cite{Panov18,Panov24}. 
The sample considered here is a freestanding (membrane) metasurface with a patterned relief.
In this approach, the phase shift of the transmitted beam is computed at varying incident light intensities $I$, allowing the nonlinear refractive index arising from the OKE to be determined, as defined by
\begin{equation}
    n = n_0 + n_2 I,
    \label{kerrdef}
\end{equation} 
where $n_0$ is the linear refractive index and $n_2$ is the nonlinear refractive index coefficient. 
This method enables spatially resolved estimation of $n_2$ across the transmitted beam profile, from which both the mean value and the standard deviation can be extracted.

The FDTD simulations are performed using the MEEP solver~\cite{OskooiRo10} to model Gaussian beam propagation through the nonlinear metasurface. For visible-wavelength simulations, the computational domain spans $4 \times 4 \times 30~\mu\mathrm{m}$ with a spatial grid resolution of $3~\mathrm{nm}$. 
The effective nonlinear refractive index coefficient of the metasurface is extracted and compared with that of bulk GaP; all values are reported as normalized, dimensionless quantities. Unless stated otherwise, the simulations are conducted at $\lambda = 532~\mathrm{nm}$. Gallium phosphide (GaP) is chosen for its high linear refractive index ($n_0 = 3.49$) and negligible extinction coefficient at this wavelength~\cite{Aspnes83}.

\begin{figure}[tbh!]
\centering
\includegraphics[width=\figsize]{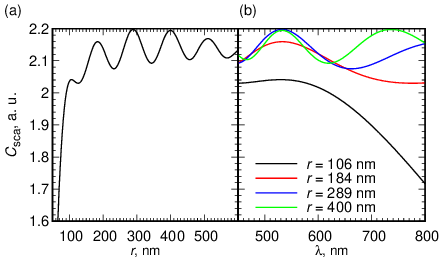}
\caption{\label{Csca_lam_r_void_GaP}
Scattering cross-sections $C_{\mathrm{sca}}$ of a spherical Mie void inside a dielectric with $n=3.49$ as functions of (a) the void radius $r$ at $\lambda = 532$~nm, and (b) the wavelength $\lambda$ for the radii that maximize the scattering in graph (a).}
\end{figure}

\section{Results and discussion}

The simplest Mie void geometry is an empty spherical cavity inside a high-index dielectric. 
The scattering cross-section $C_{\mathrm{sca}}$ of a Mie void inside GaP was calculated using \textsc{miepython}~\cite{prahl_miepython_2026} as functions of void radius and wavelength $\lambda$; the results are presented in Fig.~\ref{Csca_lam_r_void_GaP}.
As can be seen, the maxima of $C_{\mathrm{sca}}$ are only weakly pronounced.
The maxima in Fig.~\ref{Csca_lam_r_void_GaP}(a) correspond to the electric dipole ($r=184$~nm), magnetic dipole ($r=289$~nm), and electric quadrupole ($r=400$~nm) resonances.

\begin{figure}[tbh!]
{\centering
\includegraphics[width=\figsize]{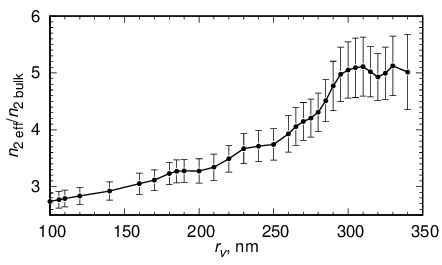}
\par} 
\caption{\label{n2_r_GaP_lattvoid}
Enhancement of the effective second-order refractive index $n_{2{\mathrm{eff}}}/n_{2{\mathrm{bulk}}}$ for a square lattice of spherical voids with lattice constant $a = 0.9$~$\mu$m in a GaP slab of thickness $h = 1$~$\mu$m plotted as a function of the void radius $r_v$.}
\end{figure} 

\begin{figure}
{\centering
\includegraphics[width=\figsize]{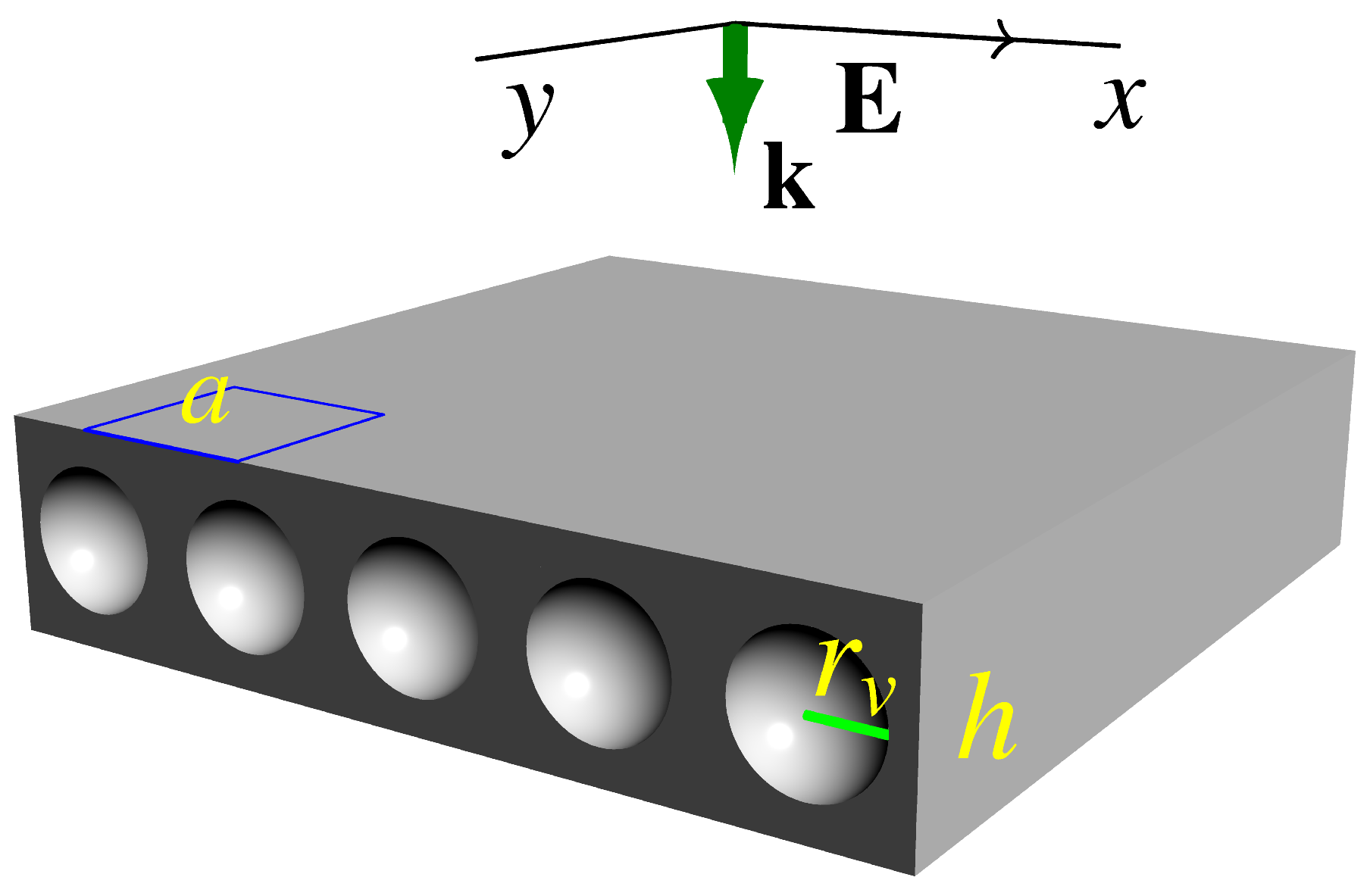} 
\par} 
\caption{\label{sphere_voids}  
Schematic of a square lattice of spherical voids in a high-index slab under normal-incident Gaussian beam illumination.}
\end{figure} 

Figure~\ref{n2_r_GaP_lattvoid} shows the dependence of the effective second-order refractive index enhancement $n_{2{\mathrm{eff}}}/n_{2{\mathrm{bulk}}}$ on the nanovoid radius $r_v$ for a square lattice of spherical voids in a GaP slab of thickness $h = 1$~$\mu$m (Fig.~\ref{sphere_voids}).
The lattice parameter is $a = 0.9$~$\mu$m, and the centers of the spherical voids are located at a depth of $h/2$.
The dependence $n_{2{\mathrm{eff}}}/n_{2{\mathrm{bulk}}}$ exhibits a gradual increase with $r_v$, reaching a maximum near $r_v = 300$~nm, which roughly corresponds to the magnetic dipole resonance.
For the opposite case of high-index spheres in vacuum, a peak-like maximum of $|n_{2{\mathrm{eff}}}|$ also occurs in the vicinity of the magnetic dipole resonance \cite{Panov19}.
For both the high-index spheres in vacuum and the inverted structure, the electric dipole resonance does not reveal any maxima of $n_{2{\mathrm{eff}}}$.
The value of $n_{2{\mathrm{eff}}}/n_{2{\mathrm{bulk}}}$ is about 5 near the maximum for the spherical voids.
In contrast, for the high-index spheres, $|n_{2{\mathrm{eff}}}|$ is up to two orders of magnitude higher than that of the bulk material \cite{Panov19}.

\begin{figure}
{\centering
\includegraphics[width=\figsize]{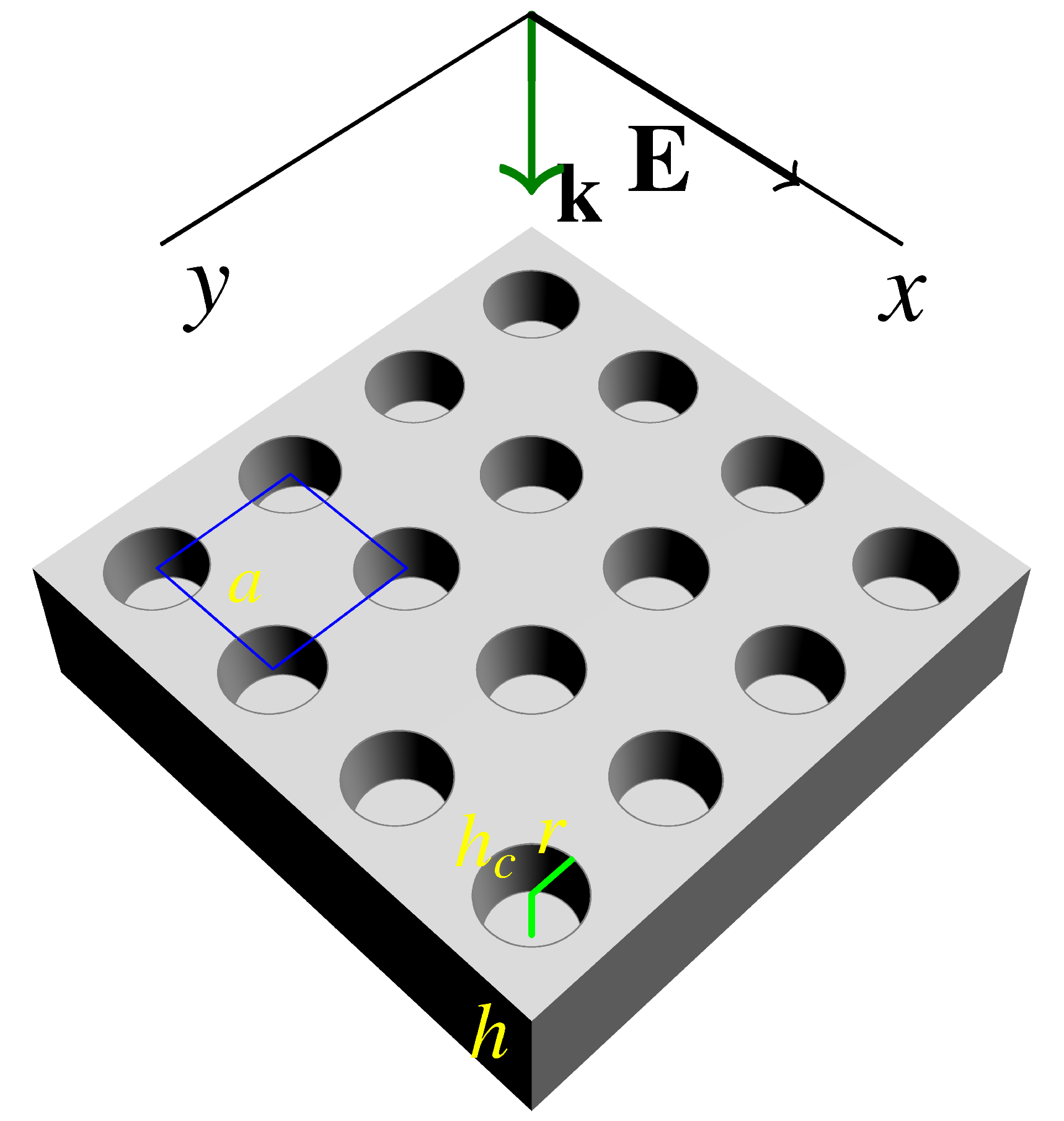} 
\par} 
\caption{\label{frustrum_voids}  
Schematic of the simulated metasurface consisting of a square lattice of truncated-cone voids in a high-index slab. A Gaussian beam is incident normally on the metasurface.}
\end{figure} 

The typical geometry of Mie voids in metasurfaces is that of truncated-cone notches on the surface.
The square lattice of such voids, as further modeled, is displayed in Fig.~\ref{frustrum_voids}.
The cones have radius $r$ at the base, which coincides with the metasurface boundary. 
The second radius $r_t$ is shorter; $h_c$ is the height of the truncated cone, and $a$ is the lattice parameter of the square lattice.
For simplicity, in this modeling $r_t = r - 30$~nm is always used.

\begin{figure}[tbh!]
{\centering
\includegraphics[width=\figsize]{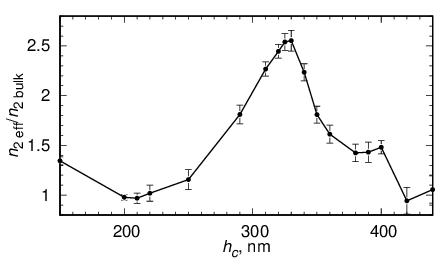}
\par} 
\caption{\label{n2_ch_GaP_lattconvoid}
Enhancement of the effective second-order refractive index $n_{2{\mathrm{eff}}}/n_{2{\mathrm{bulk}}}$ for a square lattice of truncated-cone voids with lattice parameter $a = 0.8$~$\mu$m on the surface of a GaP slab of thickness $h = 0.8$~$\mu$m plotted as a function of the cone height $h_c$.
For each $h_c$, the nanovoid base radius $r$ is chosen to maximize $n_{2{\mathrm{eff}}}/n_{2{\mathrm{bulk}}}$.}
\end{figure} 

\begin{figure}[tbh!]
{\centering
\includegraphics[width=\figsize]{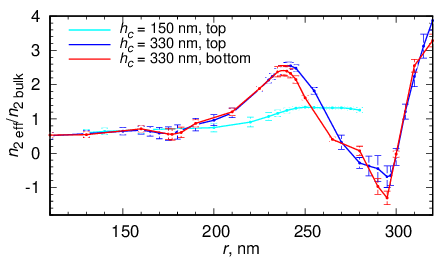}
\par} 
\caption{\label{n2_r_GaP_lattconvoid}
Enhancement of the effective second-order refractive index $n_{2{\mathrm{eff}}}/n_{2{\mathrm{bulk}}}$ for a square lattice of truncated-cone voids with lattice parameter $a = 0.8$~$\mu$m on the surface of a GaP slab of thickness $h = 0.8$~$\mu$m plotted as a function of the base radius $r$ for cone heights $h_c = 150$~nm and $330$~nm.
The cone voids with $h_c = 330$~nm are positioned on either the top or the bottom surfaces of the slab (see Fig.~\ref{frustrum_voids}).}
\end{figure} 

\begin{figure}[tbh!]
{\centering
\begin{tabbing}
\includegraphics[width=\subfigsize]{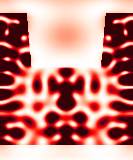}\=
\hspace{1em}\=
\includegraphics[width=\subfigsize]{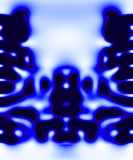}
\end{tabbing}
\par}
\caption{\label{ener_dist_GaP_lattcovoid} 
Time-averaged distributions of the electric ($\varepsilon|\mathbf{E}|^2$; left panel, red) and magnetic ($|\mathbf{H}|^2$; right panel, blue) energy densities in a GaP slab with truncated-cone voids. Parameters: lattice constant $a = 0.8$~$\mu$m, slab thickness $h = 0.8$~$\mu$m, cone height $h_c = 330$~nm, base radius $r = 241$~nm, and wavelength $\lambda = 532$~nm.}
\end{figure} 

Initially, the thickness of the GaP slab and the square-lattice parameter are fixed at 0.8~$\mu$m. 
The cone base radius and height are varied to obtain the resonant configurations.
Figure~\ref{n2_ch_GaP_lattconvoid} displays the enhancement of the effective nonlinear refractive index, $n_{2{\mathrm{eff}}}/n_{2{\mathrm{bulk}}}$, for a square lattice of truncated cones as a function of cone height $h_c$. 
The reported values correspond to radii $r$ that maximize the enhancement for each $h_c$.
Figure~\ref{n2_r_GaP_lattconvoid} illustrates the same enhancement for cone heights $h_c = 150$~nm and $330$~nm.
For the fixed slab thickness ($h = 0.8$~$\mu$m) and lattice parameter ($a = 0.8$~$\mu$m), the cone voids with $h_c = 330$~nm and $r = 241$~nm yield a maximum enhancement exceeding two.
For $h_c = 330$~nm, the enhancement increases with $r$, exhibits a shallow minimum at $r = 175$~nm, and reaches a peak at $r = 241$~nm.
At this maximum, the electromagnetic energy distribution inside the void resembles that of the electric dipole Mie resonance in a spherical void (Fig.~\ref{ener_dist_GaP_lattcovoid}).
Beyond $r = 241$~nm, the enhancement drops and subsequently rises again.
As the inter-void distance decreases, modes in the high-index regions between the pits (bridge modes) become established, as reported in Ref.~\cite{Sarbajna25}.
The case of voids positioned on the opposite face (bottom side) of the membrane metasurface for $h_c = 330$~nm is also examined in Fig.~\ref{n2_r_GaP_lattconvoid}.
The difference relative to the top-position configuration is negligible.

\begin{figure}[tbh!]
{\centering
\includegraphics[width=\figsize]{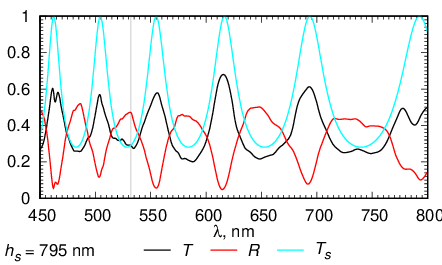}
\par} 
\caption{\label{T_lam_GaP_lattconvoid_a800_h8_r241}
Transmittance $T$ and reflectance $R$ spectra for the truncated-cone void lattice in the GaP slab with $a = 0.8$~$\mu$m, $h = 0.8$~$\mu$m, $h_c = 330$~nm, and $r = 241$~nm. 
$T_s$ is the transmittance of a plain GaP slab of thickness $h_s$, whose value is indicated below the graph.
The refractive index $n_0$ is assumed to be constant over the entire wavelength range.}
\end{figure} 

The transmittance $T$ and reflectance $R$ spectra for the truncated-cone void lattice on the surface of the GaP slab at the resonance parameters ($a = 0.8$~$\mu$m, $h = 0.8$~$\mu$m, $h_c = 330$~nm, and $r = 241$~nm) are shown in Fig.~\ref{T_lam_GaP_lattconvoid_a800_h8_r241}. 
For comparison, the transmittance $T_s$ of an unpatterned GaP slab of thickness $h_s = 795$~nm is also displayed in Fig.~\ref{T_lam_GaP_lattconvoid_a800_h8_r241}.
At this thickness, the positions of the transmittance maxima coincide with those of the patterned slab; this thickness can therefore be considered equivalent, being slightly smaller than $h = 0.8$~$\mu$m.
For $h_s = 0.8$~$\mu$m, the $T_s$ curve is shifted to longer wavelengths.
The normal-incidence transmittance is evaluated using an expression derived from the Fresnel equations (see, for example, Ref.~\cite{Bohren98}):
\begin{equation}
T_s = \frac{(1-R_f)^2}{R_f^2+1-2 R_f \cos \zeta},
\label{transmittance_slab}
\end{equation} 
where $R_f = \left|({1-n_0})/({1+n_0})\right|^2 \approx 0.31$ and $\zeta = 4 \pi n_0 h_s / \lambda$.
The slab thickness $h_s$ is adjusted to match the transmittance $T$ of the slab with nanovoids.
This unpatterned slab can be regarded as a Fabry--P\'erot cavity with reflectance $R_f$ at its facets. 
The transmittance $T$ of the nanovoid lattice in Fig.~\ref{T_lam_GaP_lattconvoid_a800_h8_r241} broadly follows the transmittance spectrum of the plain slab.

\begin{figure}[tbh!]
{\centering
\includegraphics[width=\figsize]{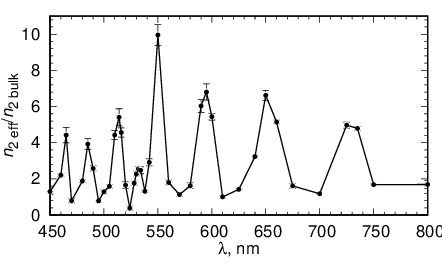}
\par} 
\caption{\label{n2_lam_GaP_25_lattconvoid_a8_h8}
Enhancement of the effective second-order refractive index for a truncated-cone nanovoid array in a GaP slab with $a = 0.8$~$\mu$m, $h = 0.8$~$\mu$m, $h_c = 330$~nm, and $r = 241$~nm plotted as a function of wavelength.
The refractive index $n_0$ is taken from Ref.~\cite{Aspnes83}.
}
\end{figure} 

Figure~\ref{n2_lam_GaP_25_lattconvoid_a8_h8} shows the spectrum of the effective second-order refractive index enhancement $n_{2{\mathrm{eff}}}/n_{2{\mathrm{bulk}}}$ for a square lattice of spherical voids in a GaP slab at the resonance parameters ($a = 0.8$~$\mu$m, $h = 0.8$~$\mu$m, $h_c = 330$~nm, and $r = 241$~nm).
As is evident, the peaks of $n_{2{\mathrm{eff}}}$, except in the vicinity of the wavelength of interest, coincide with the maxima of $R$ in Fig.~\ref{T_lam_GaP_lattconvoid_a800_h8_r241}. 
These peaks are associated with Fabry--P\'erot resonances.
The presence of the peak linked to the Mie void distorts the periodicity of the Fabry--P\'erot resonances.
The peak at $\lambda = 532$~nm is smaller than those for the Fabry--P\'erot resonances.

\begin{figure}[tbh!]
{\centering
\includegraphics[width=\figsize]{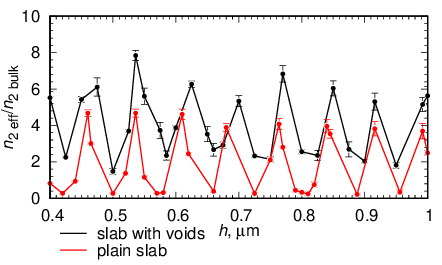}
\par} 
\caption{\label{n2_h_GaP_lattconvoid}
Enhancement of the effective second-order refractive index for a lattice of truncated-cone nanovoids in a GaP slab at $\lambda = 532$~nm plotted as a function of metasurface thickness $h$ for $a = 0.8$~$\mu$m and $h_c = 330$~nm. For each $h$, the nanovoid radius $r$ is chosen to maximize $n_{2{\mathrm{eff}}}/n_{2{\mathrm{bulk}}}$ at fixed $h_c$. The corresponding $n_{2{\mathrm{eff}}}(h)$ for the unpatterned slab is also shown.}
\end{figure} 

The dependence of $n_{2{\mathrm{eff}}}/n_{2{\mathrm{bulk}}}$ on metasurface thickness $h$ for $a = 0.8$~$\mu$m and $h_c = 330$~nm is depicted in Fig.~\ref{n2_h_GaP_lattconvoid}.
For each value of $h$, the maximum of $n_{2{\mathrm{eff}}}/n_{2{\mathrm{bulk}}}$ at fixed $h_c$ is obtained by varying $r$.
For comparison, the corresponding dependence for a plain unpatterned slab is also provided.
This slab can be treated as a Fabry--P\'erot resonator.
The periodicity of $n_{2{\mathrm{eff}}}/n_{2{\mathrm{bulk}}}(h)$ arises from the $\cos \zeta$ term in Eq.~\ref{transmittance_slab}, with a period of $h = \lambda / (2 n_0) \approx 76$~nm \cite{Panov24}.
As can be seen, the dependence of $n_{2{\mathrm{eff}}}/n_{2{\mathrm{bulk}}}(h)$ for the slab with nanovoids generally follows that for the unpatterned slab.
As with the transmittance spectrum, the maxima of $n_{2{\mathrm{eff}}}/n_{2{\mathrm{bulk}}}$ for the slab with nanovoids are slightly shifted relative to those for the unpatterned slab at $h = 0.8$~$\mu$m.
This indicates that for each thickness of the nanovoid metasurface, a slightly smaller equivalent thickness $h_s$ should be chosen.

\begin{figure}[tbh!]
{\centering
\includegraphics[width=\figsize]{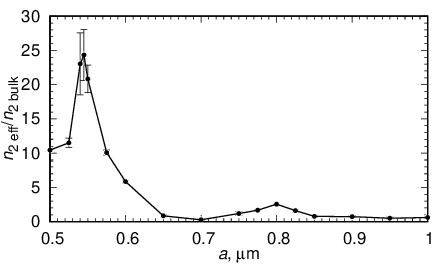}
\par} 
\caption{\label{n2_a_GaP_lattconvoid}
Enhancement of the effective second-order refractive index for a truncated-cone nanovoid array in a GaP slab at $\lambda = 532$~nm, plotted as a function of the lattice parameter $a$ for $h = 0.8$~$\mu$m, $h_c = 330$~nm, and $r = 241$~nm.}
\end{figure} 

Figure~\ref{n2_a_GaP_lattconvoid} illustrates the dependence of $n_{2{\mathrm{eff}}}/n_{2{\mathrm{bulk}}}$ on the lattice parameter for $h = 0.8$~$\mu$m, $h_c = 330$~nm, and $r = 241$~nm.
As is apparent, the global peak occurs at about $a = 0.55$~$\mu$m, where the distance between nanovoids is approximately $65$~nm.
This value is close to that reported in Ref.~\cite{Panov24} for the electric quadrupole mode in the region between hemispherical voids.
This mode, referred to as the ``bridge'' mode in Ref.~\cite{Sarbajna25}, prevails up to $a = 0.7$~$\mu$m (distance between nanovoids $\approx 220$~nm; ratio to void diameter about $1/6$).
It should be emphasized that the separation between voids in some experiments may be smaller.
The maximum at $a = 0.8$~$\mu$m, associated with the Mie void mode, is much lower.
The fixed value of $a$ for this feature suggests that this maximum is influenced by the surrounding environment.

Thus, the enhancement of the nonlinear optical effect in the metasurface with Mie voids is only severalfold and much lower (by one or two orders of magnitude) than that achievable for resonances arising in the high-index medium.
The enhancement of nonlinear optical effects is associated with the confinement of electromagnetic energy in the metasurface.
Hence, the use of Mie void metasurfaces for applications that rely on optical energy confinement is questionable.
The Mie void may, however, be useful as a reservoir for quantum emitters \cite{Fu26}, fluorophores \cite{Kuznetsov25}, or possibly quantum dots.
Otherwise, stronger electromagnetic energy confinement can be achieved in the high-index dielectric regions between the voids \cite{Panov24}.

It should also be underlined that the much larger size of Mie voids, compared to the already conventional all-dielectric nanophotonic elements, simplify their fabrication and bring them closer to macroscopic components. On the other hand, the achievable confinement of electromagnetic energy using Mie voids is significantly lower, which limits their range of applications.

\section{Conclusions}

In conclusion, the optical Kerr effect in high-index membrane metasurfaces with lattices of Mie voids has been systematically investigated using three-dimensional FDTD simulations. 
The resulting enhancement of the effective nonlinear refractive index $n_{2{\mathrm{eff}}}/n_{2{\mathrm{bulk}}}$ relative to unstructured slabs is typically below ten.

A substantially higher enhancement arises for closer placement of the voids, where a mode develops in the high-index regions between adjacent pits. 
This so-called bridge mode, associated with electric quadrupole-like or similar field distributions, prevails in GaP metasurfaces up to a separation of approximately $220$~nm between voids and can yield enhancements exceeding two orders of magnitude relative to the bulk material. 
The bridge mode thus represents a more efficient route for boosting nonlinear optical responses than the isolated Mie void resonances themselves.

The study further reveals that the nonlinear enhancement in Mie void metasurfaces is intimately linked to the spatial distribution of electromagnetic energy. 
While Mie voids confine light predominantly within low-index cavities, the accompanying field penetration into the surrounding dielectric regions is sufficient to produce measurable nonlinear effects, yet remains inferior to the confinement achievable in all-dielectric structures with high-index inclusions. 
Consequently, the practical utility of Mie void metasurfaces for applications relying solely on energy density enhancement---such as all-optical switching or intensity-dependent modulation---appears limited.

Nevertheless, the unique geometry of Mie voids offers distinct advantages in scenarios that require direct access to the confined field region. 
The open-cavity nature of the voids provides a natural host for quantum emitters, fluorophores, or nonlinear materials infiltrated into the low-index volumes, thereby enabling hybrid systems that combine strong light-matter interactions at the nanoscale with efficient outcoupling to the far field. 
Future investigations may explore such hybrid architectures, along with the extension of the Mie void concept to alternative material platforms and spectral domains.

\begin{acknowledgments}
The research was carried out within the state assignment of IACP FEB RAS (Theme FWFW-2026-0006).
The results were obtained with the use of IACP FEB RAS Shared Resource Center ``Far Eastern Computing Resource'' equipment (https://www.cc.dvo.ru).
\end{acknowledgments}

\section*{Conflict of Interest}
The author declares no conflicts of interest.

\section*{Data Availability Statement}
Data underlying the results presented in this paper are not publicly available at this time but may be obtained from the authors upon reasonable request.

\bibliography{nlphase}

\end{document}